\documentclass[aps,pra,singlecolumn,superscriptaddress, floatfix]{revtex4}
\usepackage{graphicx}
\usepackage{amssymb}
\usepackage{epsfig}
\usepackage{amsmath}
\usepackage{dcolumn}
\usepackage{bm}
\usepackage{bbm}
\usepackage{color}
\usepackage{hyperref}
\usepackage[up]{subfigure}

\newcommand{\be}{\begin{equation}}
\newcommand{\ee}{\end{equation}}
\newcommand{\bc}{\begin{center}}
\newcommand{\ec}{\end{center}}
\newcommand{\bea}{\begin{eqnarray}}
\newcommand{\eea}{\end{eqnarray}}

\begin{document}
\title{Entanglement generation in spatially separated systems using quantum walk}
\author{C. M.\ Chandrashekar}
\email{cmadaiah@phys.ucc.ie}
\affiliation{Optics  \& Quantum Information Group, The Institute of Mathematical Sciences, Tharamani, Chennai 600 113, India}
\affiliation{Department of Physics, University College Cork, Cork, Republic of Ireland}
\author{Sandeep K\ Goyal}
\affiliation{Optics  \& Quantum Information Group, The Institute of Mathematical Sciences, Tharamani, Chennai 600 113, India}
\author{Subhashish\ Banerjee}
\affiliation{Indian Institute of Technology, Jodhpur 342 011, India}

\begin{abstract}
We present a scheme for generating entanglement between two spatially separated systems from the spatial entanglement generated by the interference effect during the evolution of a single-particle quantum walk. Any two systems which can interact with the spatial modes entangled during the walk evolution can be entangled using this scheme. A notable feature is the ability to control the quantum walk dynamics and its localization at desired pair lattice sites irrespective of separation distance resulting in a substantial control and improvement in the entanglement output. Implementation schemes to entangle spatially separated atoms using quantum walk on a single atom is also presented.
\end{abstract}

\maketitle
\preprint{Version}
\section{Introduction}
\label{intro}
Entanglement is an indispensable resource for performing various quantum tasks (see \cite{HHH09} for a recent review, \cite{BKJ11} and reference therein for entanglement preparation). Several schemes have been proposed \cite{CCG99, PHB99, DLC01, BH02} for the generation and distribution of entanglement between different systems, most of which involve an initial entangling of the two systems followed by spatial separation. Such spatially separated and entangled states can be used for quantum communication protocols, for example, quantum cryptography \cite{Eke91} and quantum teleportation \cite{BBC93}. Amount of entanglement degrades with increase in spatial separation because of physical limitations and noise effect. One way of circumventing this problem would be to generate entanglement when the two systems are spatially separated. 
\par
In this article, we present a new scheme to efficiently generate entanglement 
between two spatially separated systems from a single particle system \cite{van05}. It has been shown by two of the present authors that a quantum walk evolution of a particle in a one-dimensional lattice results in the entanglement of the lattice sites after a sufficient number of walk steps \cite{GC10}. Although this spatial entanglement by itself does not have much physical significance, we nevertheless make use of it in entangling two initially unentangled systems that are spatially separated. Direct control over the quantum coin operation makes it possible to control the dynamics of the evolution of the quantum walk \cite{DM96, CSL08} which in turn allows us to optimize the entanglement output. The ability to localize the evolution at different lattice sites simultaneously, a novel phenomenon which has been discussed for the first time in this article, leads to a substantial improvement in the entanglement generated. This is a generic scheme that can be implemented to entangle any two systems that 
interact with the  modes entangled due to quantum walk. Experimental implementation of quantum walk has been reported with samples in nuclear magnetic resonance (NMR) systems \cite{DLX03, RLB05, LZZ10}; in the form of optical Galton board \cite{BMK99} and quantum quincunx \cite{DSB05}; in the continuous tunneling of light fields through waveguide lattices \cite{PLP08}; in the  phase space  of trapped  ions  \cite{SMS09, ZKG10}; with single optically trapped  atoms  \cite{KFC09}; and with single photon \cite{SCP10, BFL10}. There are various other schemes proposed to implement quantum walk in other systems \cite{RKB02, EMB05, Cha06}. Using our scheme, all these systems have the potential to generate entanglement between two spatially separated, uncorrelated systems. 
\par
This article is arranged as follows. In Section \ref{tm} we describe a toy model which has the basic ingredients of our proposal: (i) two entangled modes are generated and distributed to the distant locations of two uncorrelated systems A and B; (ii) the entanglement of these modes is then transferred to A and B via some interaction. In Section \ref{SE} we describe the discrete-time quantum walk model and the entanglement between its spatial degrees of freedom. Section  \ref{entang_gen} discusses how to use this spatial entanglement to entangle two uncorrelated, spatially separated systems A and B. The Hamiltonian modelling the interaction of these systems with the lattice sites is motivated by two examples: quantum walk with single photons and quantum walk in a spin chain; in both cases A and B are taken to be two-level systems. In Section \ref{scaling} we explain how to localize the quantum walk distribution around desired lattice sites, in such a way that the entanglement between these sites is maximized.
We then show how this affects the entanglement transferred to systems A and B. In Section \ref{imp} we propose an experimental implementations of our proposal to entangle two uncorrelated atoms in an optical lattice. We conclude in Section \ref{conc}.

\section{Toy model}
\label{tm}

\begin{figure}[ht]
\bc
\epsfig{figure=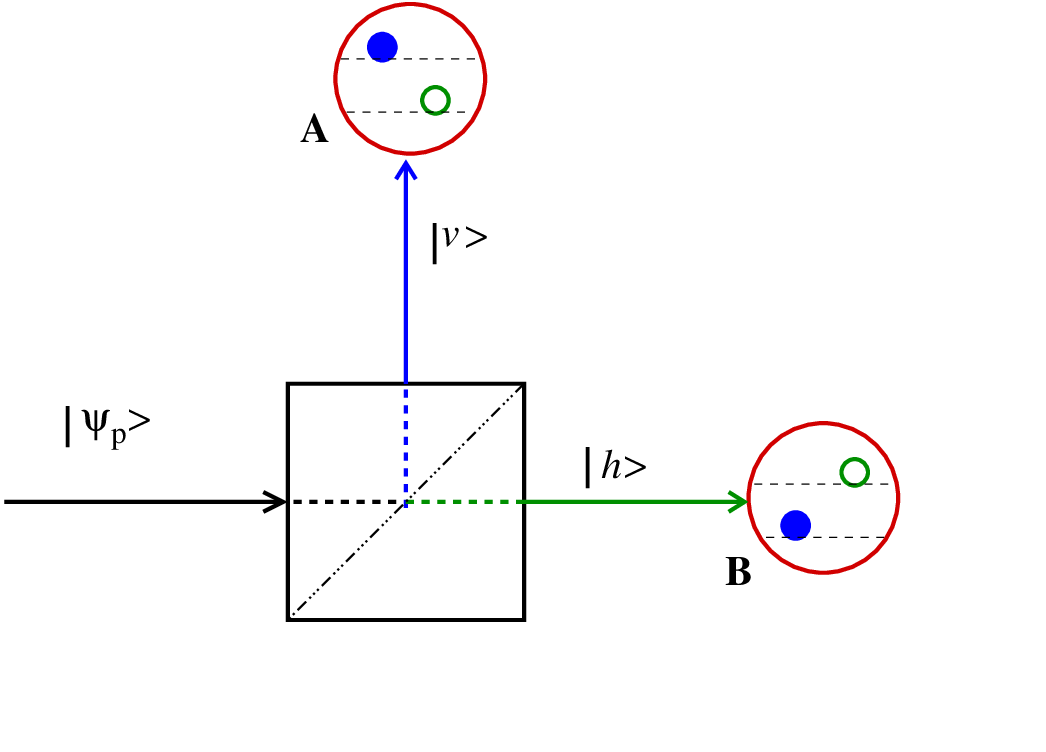, width=8.0cm}
\caption{(Color online) Photon in state $|\Psi_{p}\rangle$ when passed through the beam 
splitter gets separated making $h$ and $v$  modes which are entangled. Two, initially uncorrelated},  two-level atoms (A and B) in ground state can be entangled by interacting them with the two modes separately.
\label{bs}
\ec
\end{figure}
Before proceeding to our scheme, we will introduce the basic idea using a simple model involving a beam splitter, a photon, and two two-level atoms. The aim is to generate entanglement between the uncorrelated atoms, labeled by A and B, which are placed in distant locations (see Fig. \ref{bs}).
First, a photon in the initial state $|\Psi_{p}\rangle$ is passed through the beam splitter. The state of the photon passing through the beam splitter are spatially separated into the 
{\em horizontal} ($h$) and {\em vertical} ($v$) modes and it can be written as 
\begin{align}
|\psi_{s}\rangle = \alpha|h\rangle + \beta|v\rangle, 
\end{align}
such that $|\alpha|^2+|\beta|^2 =1$, where $|\alpha|^2$ and $|\beta|^2$
represent the probability of finding the photon in the $h$ and $v$ modes,
respectively. For convenience, we can rewrite the state of the photonic modes in
terms of the number of photons in each polarization mode:
\begin{align}
|\psi_s\rangle &=  \alpha|10\rangle_{hv} + \beta|01\rangle_{hv}.
\end{align}
The state $|10\rangle_{hv}$ ( $|01\rangle_{hv}$) represents one photon in the $h$ ($v$) mode and no photon in $v$ ($h$) mode. This state is entangled unless $\alpha$ or $\beta$ is zero. This entanglement between the polarization modes can be used to entangle A and B. 
This is done by placing atoms A and B initially in the ground state ($|g\rangle_{\rm A}$ 
and $|g\rangle_{\rm B}$) at the two exit points of the photon coming from the beam splitter. The conditions are such that if the photon is in $v$ ($h$) mode, atom A (B) will get excited, that is, $|e\rangle_{\rm A}$ ($|e\rangle_{\rm B}$). The final collective state of these two atoms can be written as:
\begin{align}
|\psi_{a}\rangle = \alpha|ge\rangle_{\rm AB} + \beta|eg\rangle_{\rm AB}.
\end{align}
This provides a very simple model of generating entanglement between two distant systems from the entanglement between the photonic modes. Its pictorial representation
is as in Fig. \ref{bs}. As we commented at the end of the Introduction, the new scheme we propose below has the ingredients of the preceding toy model; which we now describe.

\section{Spatial entanglement using single-particle quantum walk }
\label{SE}
Discrete-time quantum walk is defined on a {\it coin} Hilbert space $\mathcal
H_{c}$
and {\it position} Hilbert space $\mathcal H_{p}$. In one dimension, $\mathcal
H_{c}$ is spanned by the basis states $|0\rangle$ and $|1\rangle$ and $\mathcal
H_{p}$ is spanned by the basis state $|\psi_j\rangle$,  $j \in \mathbb{Z}$.
Each step of the quantum walk on a particle initially in superposition of the coin states at origin ($j=0$) given by, 
\begin{align}
|\Psi_{in}\rangle = \frac{1}{\sqrt{2}}[|0\rangle +
i|1\rangle] \otimes |\psi_0\rangle  \label{symmetric}
\end{align}
is implemented by applying a {\em conditional shift operation} $S$ followed by the {\em quantum coin operation} $C$. The operation $S$ can be defined such that the state
$|0\rangle$ ($|1\rangle$) moves to the left (right),  
\begin{align}
S = \sum_{j \in \mathbb{Z}} \left [ |0\rangle\langle 0| \otimes 
|\psi_{j-1}\rangle\langle \psi_{j}| + |1\rangle\langle 1| \otimes  |\psi_{j+1}\rangle\langle \psi_{j}|\right ].
\end{align}
The operation $C$  follows the operation $S$ and evolves the state at 
the new position into a superposition of the coin basis states \cite{ADZ93, DM96, ABN01}. 
It was shown in \cite{CSL08} that complete control over the dynamics of the walk can be
accomplished by choosing $C$ as a three parameter SU(2) group element; here, however, we take, for simplicity, the widely used Hadamard operator, that is,
\begin{align}
C = H = \frac{1}{\sqrt{2}}\left(\begin{array}{cc}
1&~~1\\
1& -1
\end{array}\right),
\end{align} 
which is an element of  U(2) group. 
In order to show the analytics of spatial entanglement, we consider only three steps of the walk,
after which the state of the particle can be written as:
\bea
\label{latent3}
|\Psi_{3}\rangle = S(H \otimes {\mathbbm 1} ) S (H \otimes {\mathbbm 1} ) S|\Psi_{in}\rangle =  \frac{1}
{2\sqrt{2}}[|0\rangle|\psi_{-3}\rangle \nonumber \\
 +\sqrt{3}|\chi_{-1}\rangle|
\psi_{-1}\rangle +\sqrt{3}|\chi_{+1}\rangle|\psi_{+1}\rangle +
i|1\rangle|\psi_{+3}\rangle ],
\eea
where $|\chi_{-1}\rangle = [(1+i)|0\rangle +|1\rangle]/\sqrt{3}$ and
$|\chi_{+1}\rangle = [-i|0\rangle -(1-i)|1\rangle]/\sqrt{3}$. 
Let us concentrate on the lattice sites $-1$ and $+1$ only, and denote
its position states as $|\psi_{-1}\rangle = |10\rangle$ and $|\psi_{+1}\rangle = |01\rangle$.
The reduced density matrix, after tracing out the other lattice sites and the coin degrees of freedom,
is:
\bea
\label{LatEnt}
\rho_{-1,+1}  = \frac{1}{8}[2|00\rangle\langle 00| +
  3|10\rangle\langle 10| + 3|01 \rangle\langle 01| \nonumber \\
  + 3\gamma |10\rangle\langle 01| 
   + 3\gamma^* |01\rangle\langle 10|],
\eea
where $\gamma=\langle\chi_{+1}|\chi_{-1}\rangle$ and $|00\rangle$ refers to the state when the 
walker is neither in the $+1$ nor in the $-1$ site.  The details of this trace are worked out in the appendix. The partial transpose of the above matrix will always be non-positive. Therefore,
the reduced density matrix $\rho_{-1,+1}$ represents an entangled state showing that the lattice sites $-1$ and $+1$ are entangled (cf. \cite{p96}). The next task is to use this spatial entanglement to generate entanglement between two uncorrelated systems which we will call A and B in the following.

\section{Generation of entanglement between two spatially separated systems from 
spatially entangled modes} 
\label{entang_gen}
It was shown some years ago that two distant spins A and B become entangled after
interacting with the spins of an entangled pair through a beam-splitter-like Hamiltonians \cite{LNA05}. 
The protocol, we propose, to generate entanglement between systems A and B is to : (1) evolve the desired number of quantum walk steps on a particle in lattice to generate entanglement between two spatial modes without having A and B interacting with the lattice, and (2) stop the quantum walk evolution and switch on the interaction between A and B at the desired lattice sites which are spatially entangled. The interaction of systems A and B with the entangled lattice sites depends upon the nature of the system, on which quantum walk is being performed and the properties of the system $A$ and $B$. In this section, we will consider entangling two distant spins, fermions  using quantum walk on a one-dimensional lattice consisting of spin-$\frac{1}{2}$ particles and this can be extended to bosonic system as shown in Section \ref{imp}.
\par
For  a  spin-$\frac{1}{2}$ system, the spin hops from one lattice site to another in a quantum walk evolution. By using Jordon-Wigner transformation \cite{LSM61} a spin-$\frac{1}{2}$ system can be mapped to a spinless  fermionic  system. Therefore, a one-dimensional spin-$\frac{1}{2}$ lattice with all spins but one, pointing downward can be viewed as a system consisting of a single spinless fermion. By attaching an extra coin degree of freedom with the fermion one can perform quantum walk. Now if systems $A$ and $B$ are also spin-$\frac{1}{2}$ particles, spanned by the basis $\{|g \rangle, |e \rangle\}$, interacting with the spins at $\pm l$ lattice sites, the interaction Hamiltonian can be written as 
\be
H^f_I = \hbar \omega(\sigma^+\otimes \sigma^- + \sigma^-\otimes \sigma^+).
\ee
In  the preceding Hamiltonian, $\sigma^-$ and $\sigma^+$ stands for the lowering, $|g \rangle \langle e|$ and raising, $|e \rangle \langle g|$ operator for spin respectively. 
After letting $A$ interact with spatial mode of spin at $-l$ and $B$ with spatial mode of spin at $+l$ for time $t$ given by the evolution operator 
\be
U_{I}=\exp\left(-i H^{f}_I t \right),
\ee
and the state of the system $AB$ can be written as:
\begin{align}
\rho_{AB}(t)= \sum_{ij}A_{ij}\rho_{AB}A_{ij}^{\dagger}.
\end{align}
Here $A_{ij}= \sqrt{\lambda_j}\langle i|W|\eta_j\rangle$ are Kraus operators and:
$\{|i\rangle\}$ is an orthonormal basis in $\mathcal{H}_{-l}\otimes \mathcal{H}_{+l}$, $\{\lambda_j,~|\eta_j\rangle \}_{j=1}^4$ forms the set of  eigenvalues and
eigenvectors  for $\rho_{-l,+l}$ and $\rho_{AB}$ is the initial state of the system $AB$. The unitary operator $W$, responsible for the joint evolution is:
\be
W=P[U_I\otimes U_I]P^T
\ee
where $P$ is a permutation operator such that 
\be
P[\mathcal{H}_{-l}\otimes\mathcal{H}_A\otimes\mathcal{H}_{+l}\otimes\mathcal{H}_B] = \mathcal{H}_{-l}\otimes\mathcal{H}_{+l}\otimes\mathcal{H}_A\otimes\mathcal{H}_B.
\ee
More detailed description of the process of transferring the entanglement in spin system can be seen in Ref.~\cite{CDK04} where quantum walk has been used for state transfer in a spin system.
\par
\begin{figure}[ht]
\bc
\epsfig{figure=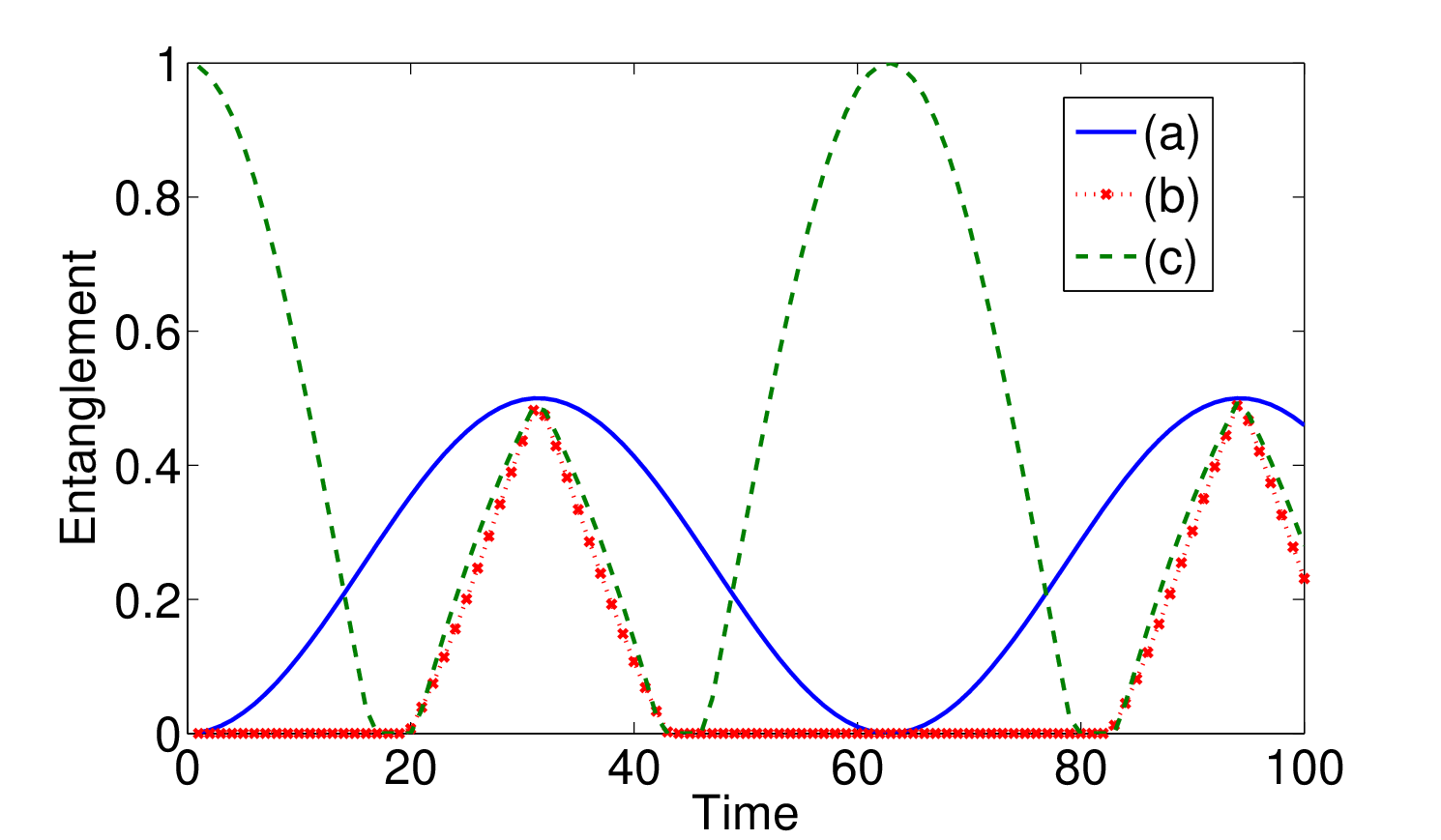, width=8.5cm}
\caption{(Color online) The evolution of generated entanglement between $A$ and $B$, spanned by the basis $\{|g \rangle, |e \rangle\}$, with different initial states under the influence of interaction with entangled lattice sites $\pm 1$ after three step of quantum walk. (a) $\rho_{AB} = |00\rangle \langle 00|$, (b) $\rho_{AB} = |11\rangle \langle 11|$ and (c) $\rho_{AB} = \frac{1}{2}(|00\rangle + |11\rangle) (\langle 00 | + \langle 11 |)$.}
\label{enttime}
\ec
\end{figure}
In Fig. \ref{enttime}, we show the evolution of entanglement between systems $A$ and $B$ for different initial states, using concurrence \cite{Woo98} as the measure. Following the previous section, we have considered the interaction of  $A$ and $B$ with the $-1$ and $+1$ lattice sites, respectively, after three steps of the walk. We observe that  the amount of entanglement depends on the initial state of the system $A$ and $B$ and different separable states achieve maximum entanglement at the same time. 

\section{Localization of quantum walk at different lattice sites}
\label{scaling}
The degree of spatial entanglement between two lattice sites depends largely on 
two points \cite{GC10} : (1) the degree of interference during the quantum walk, and (2) the value of the probability amplitude at the particular lattice sites.  Both of these can be achieved  by evolving and localizing the quantum walk around the desired lattice sites with some degree of interference. 
To further understand the contribution of localization of quantum walk to increase the degree of spatial entanglement, we will consider the lattice state after $t$ steps 
of quantum walk that  can be written as
\begin{align}
|\Psi_t\rangle_{latt} = \sum_{j} |\chi_j\rangle \otimes|\psi_j\rangle,
\end{align}
$|\psi_j\rangle$ represents a state where lattice site $j$ is occupied with other sites being empty 
and $|\chi_j\rangle$ represents the corresponding coin state. Let us say we are interested in particular lattice sites $\pm l$. Then in the reduced density matrix $\rho_{-l, +l}$, which is of the form given by Eq. (\ref{LatEnt}), with off diagonal terms $\langle\chi_{+l}|\chi_{-l}\rangle$ and diagonal terms depending on the amplitude of all $|\psi_j\rangle$ states, the states other than $|\psi_{\pm l}\rangle$ will also
contribute to $|00\rangle\langle 00|$. If the amplitude of all the $|\psi_j\rangle$ states except for $j = - l, +l$ is zero then the coefficient of $|00\rangle\langle 00|$ will be zero resulting in the maximum value of $\langle\chi_{+l}|\chi_{-l}\rangle$ which contributes for spatial entanglement.
Therefore, in localized states, that is, the states in which the amplitude is localized in a very narrow
lattice space, the coefficient of $|00\rangle\langle 00|$ will be very small resulting in maximizing  $\langle\chi_{+l}|\chi_{-l}\rangle$ and hence the amount of spatial entanglement will be more.  
\par
To realize this using quantum walk, we will begin by generalizing the previous discussions on quantum walk evolution by taking a more general coin operator
\begin{align}
\label{eq:coin}
C_{\theta} = \left(\begin{array}{cc}
\cos(\theta)&~~\sin(\theta) \\
\sin(\theta) & -\cos(\theta)
\end{array}\right).
\end{align}
During the walk evolution, if $\theta=0$ the amplitude of the two basis states move away from each other and for $\theta=\pi/2$ the amplitude shifts between the origin ($j=0$) and its neighboring positions ($\pm 1$). In both these cases the walk evolves without resulting in any interference~\cite{CSL08}. Even for $\theta$ close to $0$ and $\pi/2$ the interference effect will be very small resulting in very small or  zero spatial entanglement. When $\theta=\pi/4$ the walk evolves with good degree of interference but with a spread of the amplitudes in position space (asymptotically, the distribution is homogeneous) resulting in a very low amplitude for lattice sites $\pm l$ far away from each other.  Therefore, to maximize the amount of spatial entanglement between lattice site irrespective of the separation distance, one needs to control the quantum walk evolution in such a way that its amplitude is localized, with a good degree of interference for any lattice site separation distance.
\par
Localization of quantum walk at the origin has been discussed in Refs.~\cite{OKA05, Kon10, Cha10a, JM10, Cha11a}. Here, we briefly discuss a way to localize the walk around the desired lattice sites such that, it is scalable for site $\pm l$ far away from each other. This can be done by first delocalizing the walk to sites $\pm l$ with minimum interference resulting in large $\langle\chi_{\pm l}|\chi_{ \pm l}\rangle$ and very small $\langle\chi_{\mp l}|\chi_{ \pm l}\rangle$ followed by localization around sites $\pm l$ to improve $\langle\chi_{\mp l}|\chi_{ \pm l}\rangle$ at the cost of $\langle\chi_{\pm l}|\chi_{ \pm l}\rangle$. As discussed earlier, choosing $\theta$ very close to zero will result in two peaks  at lattice sites $\pm l = \pm t\cos(\theta)$  moving away from each other with minimal interference, where $t$ is the number of steps of the walk \cite{CSL08}. Therefore, even for very large $t$ steps, of walk delocalize at $\pm l$ position with very small $\langle\chi_{\mp l}|\chi_{ \pm l}\rangle$. 
 In Ref. \cite{Cha10a} it was shown that choosing a value of $\theta$ randomly picked from the interval $ \{\pi/4, \pi/2 \}$ at each step of the walk, localizes the quantum walk distribution around the initial position and the localization is a result of interference. Therefore, choosing $\theta$ randomly from the interval $\{\pi/4, \pi/2\}$ for each step of the after the walk is delocalized at $\pm l$ results in some improvement of $\langle\chi_{\mp l}|\chi_{ \pm l}\rangle$ at the cost of $\langle\chi_{\pm l}|\chi_{ \pm l}\rangle$ which in turn contributes for spatial entanglement. 
\begin{figure}[ht]
\bc
\epsfig{figure=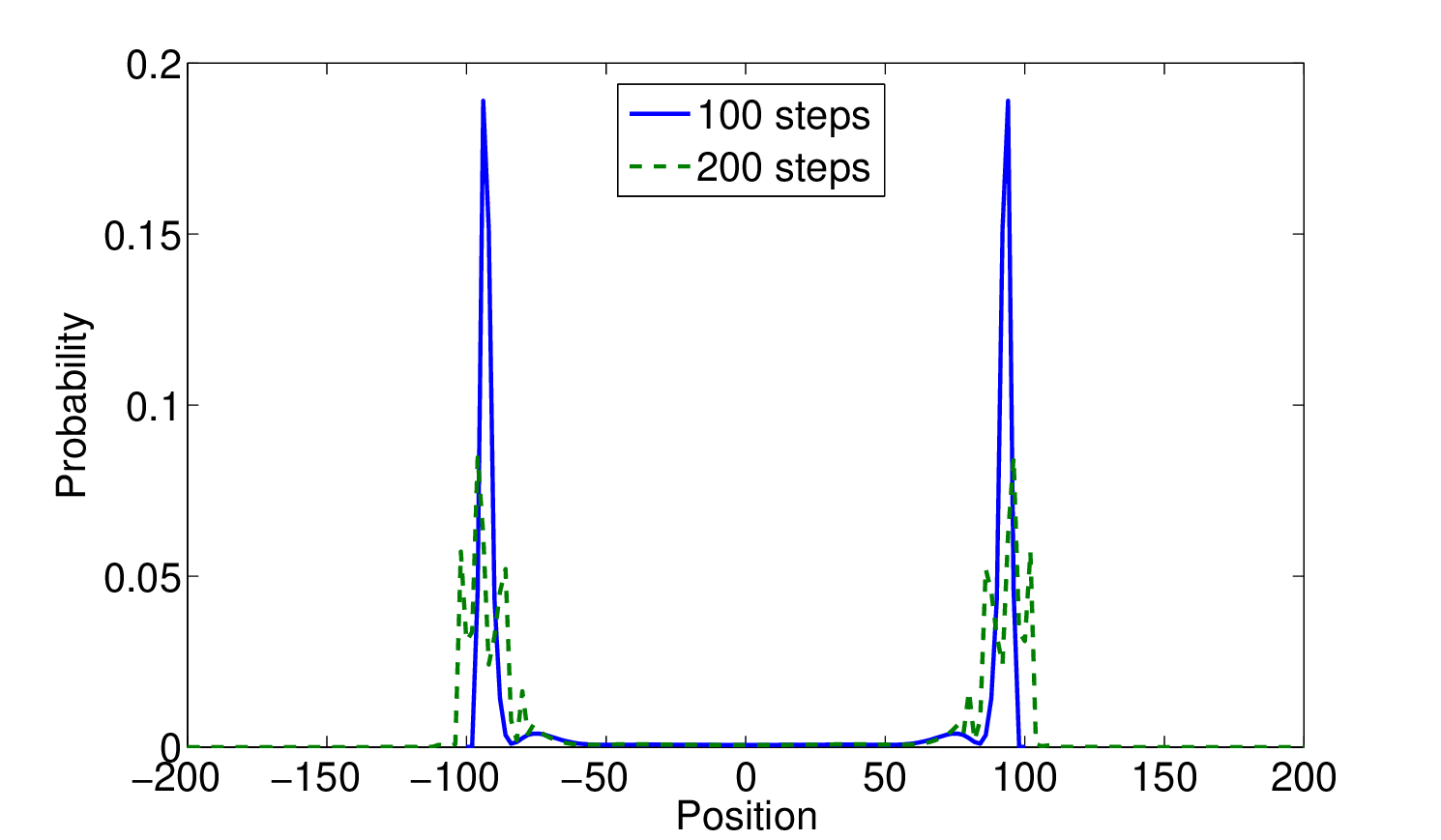, width=8.5cm}
\caption{(Color online) Localization of quantum walk around lattice sites $\pm 95$. Localization contributes to the increase in the degree of spatial entanglement which can be used to entangle two spatially separated uncorrelated systems.}
\label{egen}
\ec
\end{figure}
\par
\begin{figure}[ht]
\bc
\epsfig{figure=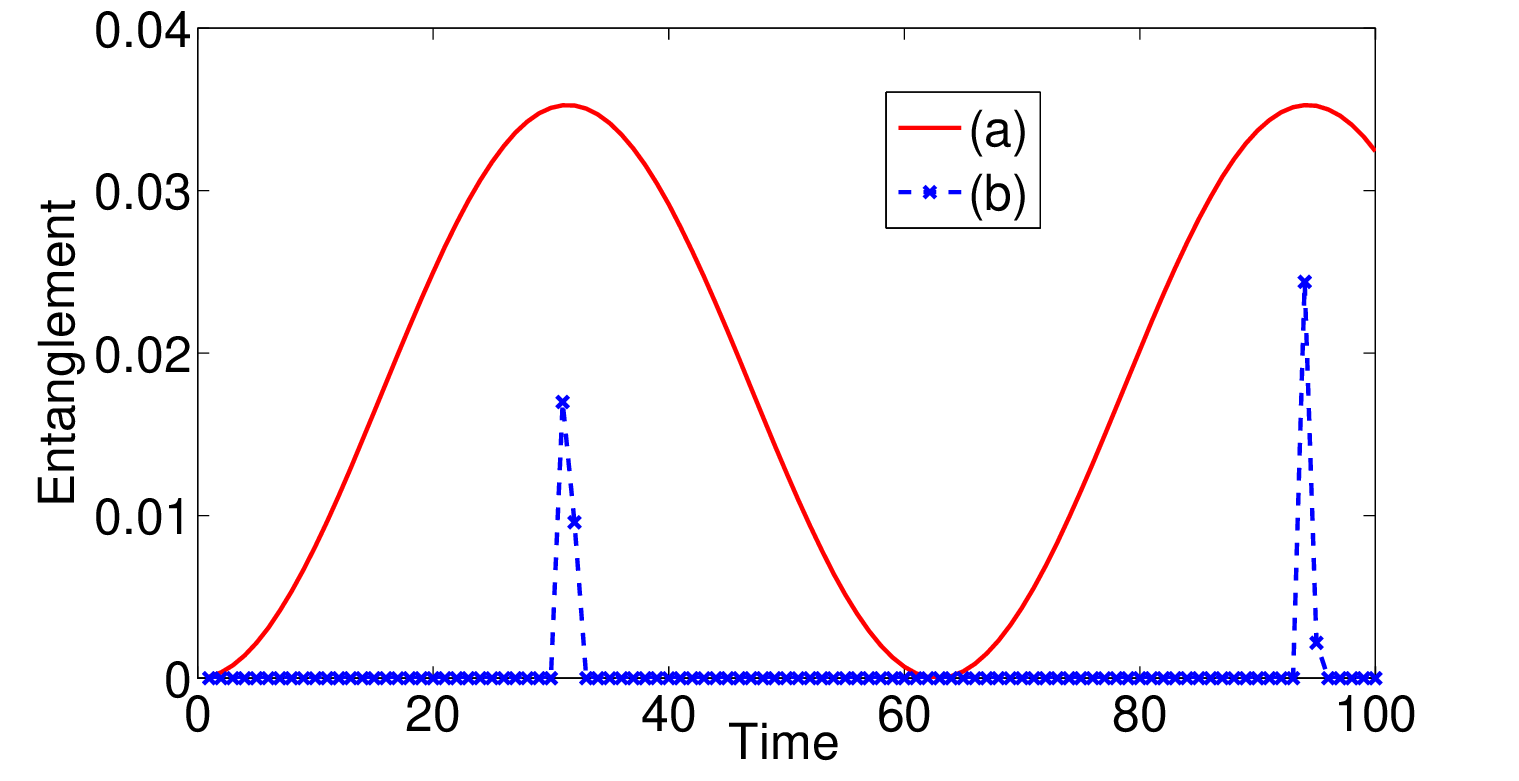, width=8.5cm}
\caption{(Color online) The evolution of generated entanglement between $A$ and $B$ with different initial states under the influence of interaction with entangled lattice sites $\pm 95$. Entanglement was generated after the walk was localized at site $\pm 95$. (a) $\rho_{AB} = |00\rangle \langle 00|$ and (b) $\rho_{AB} = |11\rangle \langle 11 |$.}
\label{enttime2}
\ec
\end{figure}

Fig. \ref{egen} shows the localization of the amplitudes at positions $-95$ and $+95$ after 100 and 200 steps of the walk; for the first 95 steps we choose $\theta \approx \pi/36$ followed by $\theta \in \{\pi/4, \pi/2\}$ for remaining steps. Note that the peaks do not move away even after 200 steps of the walk. This protocol for localization at desired sites will hold even for a very large spatial separation of lattice sites making it scalable. One should avoid using $\theta=0$ during the first part of the evolution, because as commented above, the walk, in this case, evolves without interference.  As for the entanglement transferred to the uncorrelated systems $A$ and $B$ (see the previous section), in  Fig. \ref{enttime2} we show the evolution of their concurrence when they interact with lattice sites $\pm 95$ after 200 steps of the walk. Although the amount of entanglement  in this case is smaller  than that of  Fig. \ref{enttime}, qualitatively both evolutions are identical.  The decrease in amount of 
entanglement is mainly due to the delocalization and localization of the walk which was not required for a 3-step walk used to obtain Fig. \ref{enttime}.
\section{Implementation scheme in a Physical system} 
\label{imp}

Until last year, very few experimental implementations of discrete-time quantum walk \cite{BMK99, DSB05, RLB05} were reported. However, in the  last one year,  many implementations of few steps of discrete-time quantum walk at single particle level, using trapped ions \cite{SMS09, ZKG10}; with optically trapped  atoms  \cite{KFC09}; and with photons \cite{SCP10, BFL10} have been achieved. Along with experimental implementation,  precise level of control over the single particle quantum walk evolution by controlling the quantum coin parameters have been demonstrated. These developments suggest that our scheme is directly implementable in the presently available quantum walk setups. Among the experimental setups mentioned above, we discuss a couple of them:  quantum walk with atoms and photons, in detail. For bosonic systems like atoms and photon, the interaction Hamiltonian is: 
\be
\label{interboson}
H^b_I = g(\sigma^+a + \sigma^-a^{\dagger}),
\ee
 where $\sigma^{\pm}$ are the spin raising and lowering operators and $a$, $a^{\dagger}$ are the creation and annihilation operators,  respectively. Using the time evolution operator 
 $
U_{I}=\exp\left(-i H^{b}_I t \right)$ atoms $A$ and $B$ can be entangled as discussed for the case of the spin system. 
\par
\subsection{Entangling atoms $A$ and $B$ from spatial modes of quantum walk with atom}
\label{aasystem}
\begin{figure}
\bc
\epsfig{figure=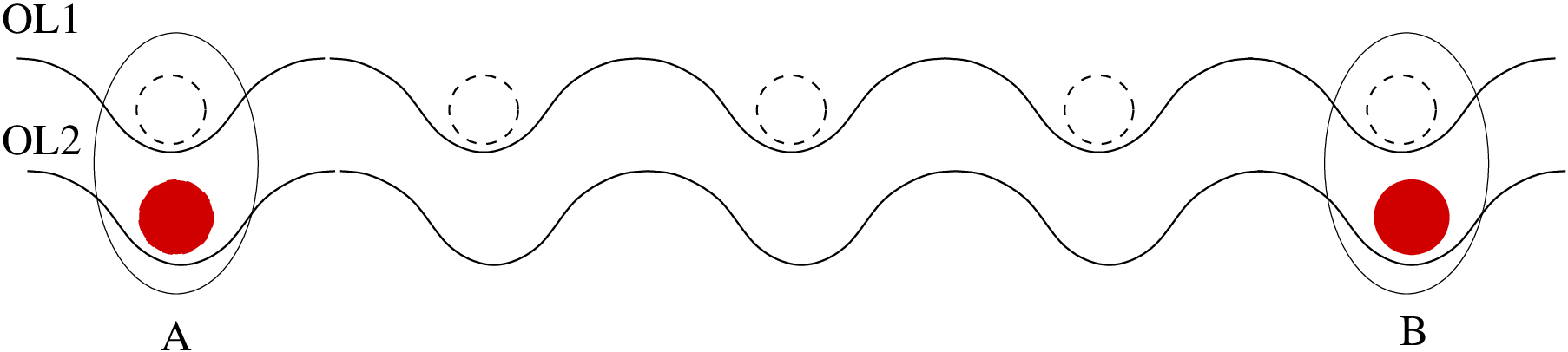, width=8.2cm}
\caption{(Color online) Entangling atoms $A$ and $B$ from the spatial modes of the quantum walk with single atom. Optical lattice (OL1) on which quantum walk is implemented is made to overlap with optical lattice on which atoms $A$ and $B$, to be entangled,  are placed (OL2).  Radio frequency pulses are used to couple states in the different lattices and entangle $A$ and $B$.}
\label{atomimp}
\ec
\end{figure}
In a recent experimental implementation of discrete-time quantum walk using neutral atoms,  single laser cooled Cesium (Cs) atoms were deterministically delocalized over the sites of a one-dimensional spin-dependent optical lattice \cite{KFC09}. Initially, the atoms distributed among the axial vibrational state were prepared in the $|0 \rangle \equiv |F = 4, m_{F} = 4\rangle$  hyperfine state by optical pumping, where $F$ is the total angular momentum, and $m_{F}$ its projection onto the quantization axis along the dipole trap axis.  A resonant
microwave radiation,   consisting of a $\pi/2$ pulse was used to coherently couples this state to the $|1 \rangle \equiv |F = 3, m_{F} = 3\rangle$ state and initialize the system in the superposition $(|0\rangle + i|1\rangle)/ \sqrt{2}\otimes |\psi_{0}\rangle$, that is,
\be
\label{atomstate}
|\Psi_{in}\rangle = \frac{1} {\sqrt{2}}(|F = 4, m_{F} = 4\rangle +  i|F = 3, m_{F} = 3\rangle) \otimes |\psi_{0}\rangle.
\ee
The state-dependent shift operation ($S$) is performed by continuous control of the trap polarization, moving the spin state $|0\rangle$  to the left and state $|1\rangle$ to the right adiabatically along the lattice axis. After $t$ steps of the quantum walk, that is, a coin operation consisting of a ($\pi/2$) pulse and a state-dependent shift operation ($S$), the 
final atomic distribution is probed by fluorescence imaging. From these images, the exact lattice site of the atom after the walk is extracted and compared to the initial position of the atom. The final probability distribution to find an atom at position $x$ after $t$ steps  is obtained from the distance each atom has walked by taking the ensemble average over several hundreds of identical realizations of the sequence.  
\par
Experimental complexity aside, the preceding protocol can be adopted to other species of ultracold atoms.  For example, in Rubidium ($^{87}$Rb) atoms in an optical trap, states $|0 \rangle \equiv |F = 1, m_{F} = 1\rangle$ and $|1 \rangle \equiv |F = 2, m_{F} = 2\rangle$ can be coherently coupled to implement quantum walk. To observe localization at a desired lattice site as discussed in Section \ref{scaling},  a $\pi/2$ pulse used as quantum coin operation during each step of the walk in the preceding protocols is replaced by a pulse with a very small value followed by a pulse randomly picked from the range $\{ \pi/4, \pi/2 \}$. The above evolution will direct the  constructive interference towards the desired lattice sites. 
\par
Once the walk is localized, the optical lattice on which the walk is implemented is made to overlap with another optical lattice on which the spatially separated atoms, 
to be entangled,  are placed. After overlapping the two lattices,  radio frequency pulses can be used to couple states in the different lattices \cite{SGC09} and entangle atoms $A$ and $B$. In Fig. \ref{atomimp}, we have shown the schematic picture of the setup. In many schemes referred in ~\cite{SGC09}, to entangle two atoms, A and B which cannot have any direct interaction between each other, they have to be made to interact via interaction Hamiltonian with two other entangled atoms. In our scheme, all we need is a  a single atoms in optical lattice and we can generate entanglement between two uncorrelated atoms in other system.

\subsection{Entangling atoms from spatial modes of quantum walk with photon}
\label{pasystem}

All three implementations of discrete-time quantum walk on photons \cite{ PLP08, SCP10, BFL10} known till date can be conveniently used to localize the quantum walk at the desired lattice site and entangle spatially separated, non-interacting atoms in an optical lattice. We consider a particular implementation using a single photon \cite{BFL10} to illustrate this further. A single photon created via parametric down conversion was initialized to a state of left-circular polarization,
\be
\label{hvstate}
|\Psi_{in}\rangle = |L\rangle = \frac{1}{\sqrt{2}}(|h\rangle + i |v\rangle) \otimes |\psi_{0}\rangle
\ee
where $|h\rangle$ and $|v\rangle$ are horizontal and vertical polarization modes.
Initialization leading to symmetric superposition state, as in Eq. (\ref{symmetric}), was done using a quarter- and a half-wave plate. Shift operator ($S$) was realized using birefringent calcite beam displacer and the coin operations used were Hadamard operation $H$ , realized using half-wave plate with axis at $\pi/8$ to the polarization plane. The above setup can be modified to localize the quantum walk at a desired lattice site by replacing the Hadamard coin operation by a more general coin operation of the form 
$C_{\theta}$ discussed in Section  \ref{scaling}.  The coin operation $C_{\theta}$ to localize the walk can be realized using a  half-wave plate with axis at $\theta/2$ to the 
polarization plane.
Once the walk is configured to localize at desired positions, the output can be directed towards spatially separated, non-interacting atoms in the optical lattice. When the photon on which the quantum walk is implemented is carefully chosen to match the transition energy level of atoms, spatially delocalized photons are absorbed by spatially separated atoms and generate entanglement between them (Eq. (\ref{interboson})). In this paper we have only discussed localization peaks which are symmetric in position space of the walk. 
As a further extension of the scheme, we can use an arbitrary SU(2) coin operation which can introduce asymmetry in the walk distribution \cite{CSL08}, and hence  localization can be realized by involving three wave plates, a half-wave plate (${\bf H}$) and two quarter-wave plates (${\bf Q}$) in any of the three configurations ${\bf Q_{1}Q_{2}H}$, ${\bf Q_{1}HQ_{2}}$ or ${\bf HQ_{1}Q_{2}}$ (concatenating the three plates) \cite{SM89} for each coin operation.
\par
Similarly, delocalized photons after the quantum walk implementation with adjustable coin operations \cite{SCP10} and quantum walk with photons in a waveguide lattice can be directed towards atoms in a spatially separated optical lattice and entangle them.

\section{Conclusion} 
\label{conc}
We have presented a scheme for generating entanglement between two distantly located uncorrelated systems using a single particle quantum walk. Spatial modes entangled, due to the interference effect from the quantum walk evolution, were transferred to two distantly located systems and  effectively used for the purpose of entangling them. By controlling the amount of entanglement generated between two spatial modes of the quantum walk system we have also shown that the entanglement generated between the two initially uncorrelated systems can be controlled.  The control is brought about by quantum coin operations, localization of the walk evolution at different lattice sites. This scheme to entangle two systems using single particle quantum walk can be directly scaled to simultaneously entangle large systems using many particle walk. We have shown that all the elements of the present scheme are within reach of current experimental setups on which quantum walks have been implemented \cite{RLB05, LZZ10, PLP08, 
SMS09, ZKG10, KFC09, SCP10,BFL10}. This scheme can come in handy for systems in which only selected sites can be accessed for inducing interaction with uncorrelated systems. Limitations of the resource and the environmental effect is an issue at this point of time, however, experimental progress accounting for decoherence \cite{BFL10} is a positive thrust in the direction of scaling the implementable steps of the walk.\\
\\
\appendix
\section{Obtaining Eq. (\ref{LatEnt}) from  Eq. (\ref{latent3})}
\label{Appendix}
The state of a particle on lattice site $l$ is represented by $|\psi_l\rangle$
which can equivalently be represented by a creation operator
$a_l^{\dagger}$ acting on the lattice site $l$, i.e,
$a^{\dagger}_l|\Omega\rangle$. Here $|\Omega\rangle$ represents the
vacuum state of the system. In the case of spin half lattice, this vacuum
can be the states in which all the spins are down and the creation
operator is the operator which maps down spins to up, i.e,
$a^{\dagger}|\downarrow\rangle = |\uparrow\rangle$. Using this
notation one can rewrite the equation (7) as:
\begin{eqnarray}
|\Psi_3\rangle_{latt} =& \frac{1}{2\sqrt{2}}\left[|0\rangle
  a_{-3}^{\dagger}+\sqrt{3}|\chi_{-1}\rangle a_{-1}^{\dagger}\right.\nonumber\\
 &\left.+ \sqrt{3}|\chi_1\rangle a_1^{\dagger} + i|1\rangle
  a_3^{\dagger}\right]|\Omega\rangle. 
\end{eqnarray}
Since we are interested only in lattice sites $-3,~-1,~+1,~+3$, the
state $|\Omega\rangle = |0000\rangle$. From here we can write the
state $\rho_{-1,+1}$ for the lattice site $-1$ and $+1$. To get that
first we need to trace out the coin degree of freedom. So the state of
the particle $\rho_p$ after tracing out the coin degree of freedom is:
\begin{eqnarray}
\rho_p &=& {\rm Tr}_c(|\Psi_3\rangle\langle \Psi_3|) \nonumber \\
&=& \frac{3}{8} \left(|\psi_{+1}\rangle\langle\psi_{+1}| +
|\psi_{-1}\rangle\langle\psi_{-1}| + \gamma|\psi_{-1}\rangle\langle\psi_{+1}| 
+ \gamma^*|\psi_{+1}\rangle\langle\psi_{-1}|\right) \nonumber \\
& & + \frac{1}{8}\left( |\psi_{-3}\rangle\langle\psi_{+3}| +
|\psi_{+3}\rangle\langle\psi_{-3}|\right)\nonumber\\
&& + \frac{1}{8} \left[
(1-i)|\psi_{-3}\rangle\langle\psi_{-1}| + i|\psi_{-3}\rangle\langle\psi_{+1}|
+ i|\psi_{+3}\rangle\langle\psi_{-1}| -
(1-i)|\psi_{+3}\rangle\langle\psi_{+1}| + H.c.\right]\nonumber\\
&&+\frac{1}{8}\left(|\psi_{-3}\rangle\langle \psi_{-3}| +
|\psi_{+3}\rangle\langle \psi_{+3}|\right). 
\end{eqnarray}
In our new notation the above equation can be written as:
\begin{eqnarray}
\rho_p &=&\frac{3}{8} \left(a_{+1}^{\dagger}|\Omega\rangle\langle\Omega|a_{+1} +
a_{-1}^{\dagger}|\Omega\rangle\langle\Omega|a_{-1} + 
\gamma a_{-1}^{\dagger}|\Omega\rangle\langle\Omega|a_{+1}
+ \gamma^*a_{+1}^{\dagger}|\Omega\rangle\langle\Omega|a_{-1}\right) \nonumber\\
&&+\frac{1}{8}\left( a_{-3}^{\dagger}|\Omega\rangle\langle\Omega|a_{+3} +
a_{+3}^{\dagger}|\Omega\rangle\langle\Omega|a_{-3}\right) \nonumber\\
&&+ \frac{1}{8} \left[
(1-i)a_{-3}^{\dagger}|\Omega\rangle\langle\Omega|a_{-1} + 
ia_{-3}^{\dagger}|\Omega\rangle\langle\Omega|a_{+1}
+ ia_{+3}^{\dagger}|\Omega\rangle\langle\Omega|a_{-1} -
(1-i)a_{+3}^{\dagger}|\Omega\rangle\langle\Omega|a_{+1} + H.c.\right]\nonumber\\
&&+\frac{1}{8}\left(a^{\dagger}_{-3}|\Omega\rangle\langle \Omega|a_{-3} +
a^{\dagger}_{+3}|\Omega\rangle\langle \Omega|a_{+3}\right)
\end{eqnarray}
which can be written as:
\begin{eqnarray}
\rho_p &=&\frac{3}{8} \left(|0010\rangle\langle 0010|
+|0100\rangle\langle 0100| + \gamma |0100\rangle\langle 0010|+
\gamma^*|0010\rangle\langle 0100|\right) \nonumber \\
&&+\frac{1}{8}\left(|1000\rangle\langle 0001| +|0001\rangle\langle 1000|\right) \nonumber\\
&&+
\frac{1}{8} \left[(1-i)|1000\rangle\langle 0100| +i|1000\rangle\langle
  0010|+ i|0001\rangle\langle 0100| -(1-i)|0001\rangle\langle 0010|+
  H.c.\right] \nonumber\\
&&+\frac{1}{8}\left(|1000\rangle\langle 1000| +
|0001\rangle\langle 0001|\right).
\end{eqnarray}
From here we can obtain Eq. (\ref{LatEnt}) by tracing out the sites $-3$ and
$+3$,
\begin{align}
\rho_{\pm 1} = {\rm Tr}_{\pm 3}(\rho_p) = _{\pm 3}\langle 00|\rho_p|00\rangle_{\pm 3}+ 
_{\pm 3}\langle 01|\rho_p|01\rangle_{\pm 3}+ 
_{\pm 3}\langle 10|\rho_p|10\rangle_{\pm 3}+ 
_{\pm 3}\langle 11|\rho_p|11\rangle_{\pm 3}\\
\rho_{\pm 1} =\frac{3}{8}\left(|01\rangle\langle 01| + |10\rangle\langle 10| +
\gamma |10\rangle\langle 01| + \gamma^*|01\rangle\langle 10|\right) +
\frac{1}{4}\left(|00\rangle\langle 00|\right).
\end{align}
In the preceding expression $|00\rangle_{\pm 3}$ refers to the state when the walker is neither in the $+3$ nor in the $-3$ state.


\end{document}